\documentclass[preprint,aps,showpacs]{revtex4}
\usepackage{graphics,graphicx,epsfig}

\begin{document}
\title{Scaling of island size and capture zone distributions in submonolayer growth}
\author{T. J. Oliveira${}^{1,2,(a)}$ and F. D. A. Aar\~ao Reis${}^{1,(b)}$
\footnote{a) Email address: tiago@ufv.br\\
b) Email address: reis@if.uff.br}}
\address{${}^{1}$ Instituto de F\'\i sica, Universidade Federal
Fluminense, Avenida Litor\^anea s/n, 24210-340 Niter\'oi RJ, Brazil\\
${}^{2}$ Departamento de F\'isica, Universidade Federal de Vi\c cosa, 36570-000, Vi\c cosa, MG, Brazil}

\date{\today}

\begin{abstract}

Island size and capture zone distributions (ISD, CZD) are studied numerically in submonolayer growth
with various critical island sizes and shapes.
CZD scaled by the variance show excellent agreement with the Wigner surmise, confirming the
Pimpinelli-Einstein approach for large CZs / large island dynamics.
The ISD decay as $\exp{\left( -s^\gamma\right)}$, with $\gamma =4$, $\approx 2.4$ and $2$ for point,
fractal, and square islands, respectively. A scaling approach explains the values of $\gamma$ from the
Gaussian decay of CZD and the efficiency of islands to capture diffusing adatoms. 

\end{abstract}
\pacs{68.43.Hn, 68.35.Fx, 81.15.Aa, 05.40.-a}

\maketitle

Since island nucleation in the submonolayer regime determines several features of the
subsequent growth of a thin film \cite{etb,pimpinelli,venables,ratsch}, it motivated many
theoretical works to explain the island size distributions (ISD) and the capture zone distributions (CZD).
Theoretical distributions are frequently compared to experimental data
from various materials and processes or with simulation data from atomistic models,
providing estimates of parameters such as critical nucleus sizes and binding energies.
Recent works on submonolayer growth of organic molecules \cite{mulheranPRL2008,clancy2006,zhang}
and colloidal epitaxy \cite{ganapathy} increase the interest in the subject
for the possibility of extending the knowlegde on atomic epitaxy.

However, the exact forms of the CZD and ISD still remain unknown and, consequently, the basic
mechanisms governing their kinetics are unclear.
Most theoretical approaches assume irreversible aggregation of adatoms to islands whose
size exceeds a critical value $i$. Important examples are that of Amar and Family (AF),
who proposed an empirical formula for
the ISD \cite{af1995}, the Mulheran and Blackman (MB) approach to relate ISD to
distributions of areas of Voronoi polygons \cite{mb}, and the recent proposal
of Pimpinelli and Einstein (PE)  \cite{pe} that CZD are described by the Wigner surmise (WS)
from random matrix theory \cite{rmt}.
The AF formula is widely used to fit the peaks of experimental
ISD and to estimate critical island sizes (see e. g. Ref. \protect\cite{ruiz}), but deviations
from point island model distributions are significant \cite{etb,afmatres}.
The solution of equations from the MB approach provides ISD very close to point
island model simulations, but it is not so popular as the AF formula for fitting experimental data.
The WS works better than the MB curves for fitting simulated CZD of point and 
circular islands \cite{pe}, but recent numerical work by 
Amar and Evans groups \cite{shi,li} showed deviations in the peaks and in the left tails.
Alternatively, a Gamma distribution was used to fit CZD \cite{fgamma}.

Here we will analyze very accurate simulation data of ISD and CZD of 
growth models of point and extended (fractal and square) islands with irreversible aggregation
($i=1$ and $2$). The CZD are fitted by the WS after rescaling by the variance,
which reduces deviations in the peaks and left tails and hightlights the
Gaussian right tail predicted by PE for all island shapes. The ISD shows deviations from the AF formula,
but have universal right tail decays dependent on the island shape and related
to the Gaussian CZD. This confirms the PE approach for describing the
dynamics of large islands and large capture zones (CZs).

We performed simulations in square lattices of size $2048\times 2048$, and confirm
negligible finite-size effects with runs in size $1024\times 1024$. Diffusion-to-deposition
ratios $R\equiv D/F$ between ${10}^6$ and ${10}^{10}$ were analyzed, with critical island
sizes $i=1$ and $i=2$. For $i=2$, detachment probability $\epsilon={10}^{-3}$ was used with point
islands to facilitate island growth and reduce fluctuations, but for extended islands we
used $\epsilon =1$. Averages are taken from ${10}^4$ lattice configurations for each
$i$, $R$ and coverage $\theta$. In point island models, all atoms of an island aggregate
at a single lattice site. The model version of Shi, Shim, and Amar \cite{shi,shiPRB2005} was
simulated, with an adatom aggregating to an island as it is deposited or
hops onto it (however, simulations of the neighbor-attachment version \cite{eb} leads to similar
conclusions). In fractal island models, each atom permanently aggregates
at the site where it collides with an island \cite{jensen}, generating branched clusters that
resemble those of diffusion-limited aggregation (DLA) \cite{witten}. In the square island model,
instantaneous relaxation to compact (square) shape is perfomed after an adatom collides with
an island \cite{sqevans}. A site belongs to the CZ of an island if its distance to the island
border is smaller than its distance to any other island border. The CZ area $x$
is the number of sites in the CZ, and the island size $s$ is its number of atoms.

It is expected that the probability density of CZ area $x$ follows the scaling form
\begin{equation}
P(x)=\frac{1}{\langle x\rangle}
f{\left( \frac{x}{\langle x\rangle}\right)} ,
\label{scalingtrad}
\end{equation}
and an equivalent relation is assumed for the density of islands of size $s$, $Q(s)$. 
The alternative scaling of CZD with the variance
$\sigma_x\equiv {\langle{{\left( x-{\langle x\rangle}\right)}^2}\rangle}^{1/2}$ as
\begin{equation}
P(x) = \frac{1}{\sigma_x} g{\left( u\right)} \qquad ,\qquad  u\equiv \frac{x-\overline{x}}{\sigma_x} ,
\label{scalingP}
\end{equation}
was used in a recent study of roughness distributions \cite{intrinsic} to
reduce the effect of scaling corrections in $\overline{x}$ and 
$\sigma_x$ (here, corrections are expected to arise from small CZs, whose dynamics continuous
theories fail to describe).
For comparison, the WS is
\begin{equation}
P_\beta(x) = a_\beta x^{\beta} \exp{\left( -b_\beta x^2\right)} ,
\label{ws}
\end{equation}
with the parameter $\beta = \frac{2}{d}\left( i+1\right)$ predicted by the
PE theory, where $d$ is the substrate dimension ($d=2$ in
this work). The AF formula for ISD is
\begin{equation}
f_i\left( u\right) = C_i u^i \exp{\left( -ia_i u^{1/a_i}\right)},
\label{af}
\end{equation}
where $u\equiv s/\langle s\rangle$, $C_i$ and $a_i$ are normalization constants.

In Fig. 1a we show scaled CZD for point islands with $i=1$. An excellent collapse of the data for
different $R$ is observed with scaling by the variance, as shown in the inset of Fig. 1a. This is
an important starting point for comparison with ($R$-independent) theoretical approaches.
This contrasts to the scaling by the average, which shows significant $R$-dependence of the peak
heights \cite{shi}. Fig. 1a also shows four theoretical curves for comparison with the simulation
data. The WS with $\beta=2$, predicted by PE theory, is a good fit for the peaks
under this rescaling (inset of Fig. 1a)), but shows deviations in the tails, as previously observed
in Refs. \protect\cite{shi,li}. The Gamma distribution that best fits the CZD peaks
has remarkable differences in both tails. The WS with $\beta=3$ was suggested in Refs.
\protect\cite{li,pereply} and turns out to be a good fit for the peak and left tail of the simulation data,
but also differs from the right one. Finally, the generalized Gamma (GG) fit
$g(x)\sim x^\beta \exp{\left( -bs^n\right)}$ with $\beta=4$ and $n=1.5$, proposed in Ref. \protect\cite{li},
has small deviation only in the right tail.
 
Fig. 1b shows the same simulation data of Fig. 1a with abscissa $u^2$. The fit for $u^2 \geq 15$
highlights the Gaussian right tail. This is the universal decay of CZD predicted by PE theory.

Fig. 1c shows scaled CZD for point islands with $i=2$, compared to the WS with $\beta=3$, as predicted
by PE theory. Good data collapse is also obtained with this rescaling (inset of Fig. 1c).
There is discrepancy from the PE curve in the left tail, but the agreement in the right tail is
very good, which confirms the Gaussian decay.

The comparison of the skewnesses ${\langle {\left( x-\langle x\rangle \right)}^3\rangle}/\sigma^3$
of the simulated CZD and the WS gives additional support to the PE approach:
simulation gives $S=0.51\pm 0.01$ and $S=0.41\pm 0.01$ for $i=1$ and $i=2$,
respectively, while the values of the WS are $0.486$ for $\beta=2$ and $0.406$ for $\beta=3$.
It is also observed that CZD for point islands have a negligible dependence 
on the substrate coverage $\theta$ when scaled by the variance.

Scaled CZD of fractal islands with $i=1$ are shown in Fig. 2a, with the predicted
WS curve ($\beta=2$). The good fit in the tails confirms the Gaussian decay, and scaling by
the variance also leads to a good data collapse for different $R$ and agreement with the WS peaks. 
Results for square islands with $i=2$ are shown in Fig. 2b, again in good agreement with
the respective WS curve ($\beta=3$). Similar good fits by WS curves are obtained for fractal islands
with $i=2$ and square islands with $i=1$.

For fractal islands, the good data collapse shown in Fig. 2a is obtained for coverages
$\theta \lesssim 0.20$. For square islands, sub-island overlap enhances coalescence effects,
thus the behavior shown in Fig. 2b is observed only for $\theta \lesssim 0.10$.
Also note that, for small $R$ (typically $R\lesssim 10^{5}$), the right tails for extended islands
deviate to simple exponential decays, and data for $R={10}^6$ show crossover behavior.

These results are striking evidence that the PE theory actually captures the main ingredients of the
dynamics of large CZs. Indeed, the phenomenological model introduced by PE \cite{pe}
extracts the Wigner distribution from a Langevin equation, which is a continuous
approach expected to apply for large $x$. The deviations in scaling by the average
(Eq. \ref{scalingtrad}) and the better data collapse in scaling by
the variance (Eq. \ref{scalingP}) confirm that scaling corrections are reduced by the latter.
Those corrections are related to small CZs, whose dynamics is
not accurately described by the continuous PE approach. For instance, this
explains the larger deviations for point islands with $i=1$,
particularly in the left tails (as previously observed in Refs. \protect\cite{shi,li}):
since adatoms have to move for long times before reaching an existing island,
the probability of two diffusing adatoms to meet (nucleating a new stable island) is large,
leading to division of large CZs into smaller ones.
For $i=2$, the instability of two-atom islands reduce this effect.

The scaling by the variance makes the peaks of CZDs indistinguishable for different i?s in a linear plot, in contrast to the log-linear plots that highlight the differences in the tails. Thus, it will probably be not useful to distinguish the critical nucleus in experimental data, where ?uctuations in the tails are large. However, it is certainly important for comparisons in theoretical works.

Now we present results for ISD.

Fig. 3a shows results for point islands. The variable ${\left( s/\langle s\rangle\right) }^4$ highlights the
right tail decay as $\exp{\left( -s^4\right)}$ for large $R$. The inset of Fig. 3a shows the
same data scaled by the variance $\sigma_s$ [with $z \equiv (s -\langle s\rangle)/\sigma_{s}$].
This rescaling does not provide collapse of ISD data for different
$R$ and $\theta$ and introduces corrections in the right tail decay shown in the main plot.
Fig. 3a also shows the AF formula, which does not represent these ISD because
it gives $Q(s)\to 0$ as ${s/\langle s\rangle}\to 0$, as noted before \cite{etb,afmatres}.

For fractal islands, scaling by the variance provides good data collapse for different $R$,
as shown in Fig. 3b.  The AF formula is a good fit for the ISD peaks (inset of Fig. 3b),
which explains its wide use, but it does not represent the right tail (main plot).
The decay of the ISD is only slightly faster than the Gaussian decay of the CZD.
The good fits of the right tails by $\exp{\left( -z^{2.4}\right)}$ (main plot)
or $\exp{\left( -s^{2.4}\right)}$ (if scaled by the average) explain the choice of the abscissa in Fig. 3b.

The ISD for square islands are shown in Fig. 3c. Good data collapse for different $R$ is
obtained in the right tails with scaling by the variance, despite the large differences in the
left tails and peak heights. The same Gaussian decay of the CZD is obtained. 
Again, the AF formula shows deviations from the simulated ISD.

The ISD for extended islands (fractal or square) have significant coverage dependence.
For very low coverage, point island decay [$\sim e^{\left( -s^{4}\right)}$] is found,
and, for large coverage, the right tails tend to simple exponentials. The behavior
shown in Figs. 3a-c is typical of a narrow (though widely studied) range
$0.05 \leq\theta\leq 0.2$. Analogous results are obtained for $i=2$.

Now we establish a connection between the CZD and the ISD which explains the observed
decays of the latter. Since the decay of CZD is universal, as predicted
by PE, it is natural to expect that the decay of ISD will be related to that Gaussian
decay and to the particular island shape. Working with large islands and capture zones
is certainly necessary for such continuous approach.

For fractal islands, it is plausible that larger islands
have larger CZ. This is shown for circular islands in Ref. \protect\cite{mulheranrobbie}.
Thus, consider that islands of size $s$ have CZ of typical size $x$, so that $P(x)\sim Q(s)$.
The shape of that island resembles that of a DLA cluster for small coverage because most
of the attached atoms came from the island neighborhood by diffusion, and not by deposition
inside that island. Thus, the typical radius of the island is $R\sim s^{1/D_F}$, with
the fractal dimension $D_F\approx 1.694$ of DLA \cite{meakin83}. The area
of the CZ is expected to scale as the area of a circle with that radius, thus
$x\sim R^2\sim s^{2/D_F}$. Consequently, the Gaussian tail of the CZD implies a tail for the ISD
as $\exp{\left( -x^2\right)}\sim \exp{\left( -s^{4/D_F}\right)}
\approx \exp{\left( -s^{2.36}\right)}$. This faster-than-Gaussian decay is in excellent
agreement with the results in Fig. 3b.

For compact (square or circular) islands, the fractal dimension is $D_{F}=2$,
thus the same arguments lead to a Gaussian decay of the ISD,
again in excellent agreement with our numerical results (Fig. 3c).

For point islands, consider a typical CZ of area $x$ of an island with size $s$ ($s$ atoms
in a single site),  with $x,s\gg 1$. The radius of this CZ is $R\sim x^{1/2}$, and its border
has length of order $R$. It is surrounded by other CZs, whose number grows proportional to
the border length $R$ (typically they are small CZs neighboring a large one)..
Now consider islands whose CZ is of size $x+\Delta x$ and size $s+\Delta s$.
The increase in the size is due to atoms deposited in the
additional area $\Delta x$ which were able to reach this point island, instead of being
absorbed by the neighboring islands. We expect that a fraction $\sim \frac{1}{R}$ of those
deposited atoms will actually reach the central point island.
Thus, $\Delta s\sim \Delta x \frac{1}{R}$, which gives $\frac{ds}{dx}\sim
\frac{1}{x^{1/2}}$ in the continuum limit and, consequently, $x\sim s^2$
(this means that large CZ area $x$ corresponds to not so large island size $s\sim x^{1/2}$).
The Gaussian tail of the CZD implies a tail for the ISD as
$\exp{\left( -x^2\right)}\sim \exp{\left( -s^4\right)}$.
This is also in excellent agreement with our numerical results for $i=1$ (Fig. 3a) and
reasonable agreement for $i=2$.

In summary, we studied numerically the ISD and CZD in submonolayer growth of islands
with different shapes (point, fractal, and square) and critical sizes $i=1$ and $i=2$.
The Wigner surmise describes the CZD when scaled by the variance of CZ area, with
deviations in the left tails for point islands with $i=1$. This confirms the superuniversal Gaussian
decay predicted by Pimpinelli and Einstein, in contrast with the fits proposed by other
approaches, and improves recent numerical work which did not focus
on the CZD decay and used scaling by the average. The ISD are also not represented by
well known fitting formulas, but universal decays of the right tails are found
for each island shape and are explained by connections to the Gaussian
tails of the CZD. We believe that these results motivate experimental efforts to
measure accurate ISD and CZD, which are possible with the advance in microscopy
techniques \cite{rost}. It also motivates theoretical efforts to understand
ISD features other than the right tails, the
reversible growth processes of atomic islands, and growth of more complex
nanostructures \cite{ganapathy}.

\acknowledgments

The authors acknowledge support from CNPq and FAPERJ (Brazilian agencies).


\newpage

\begin{figure}[!h]
\includegraphics[width=8cm]{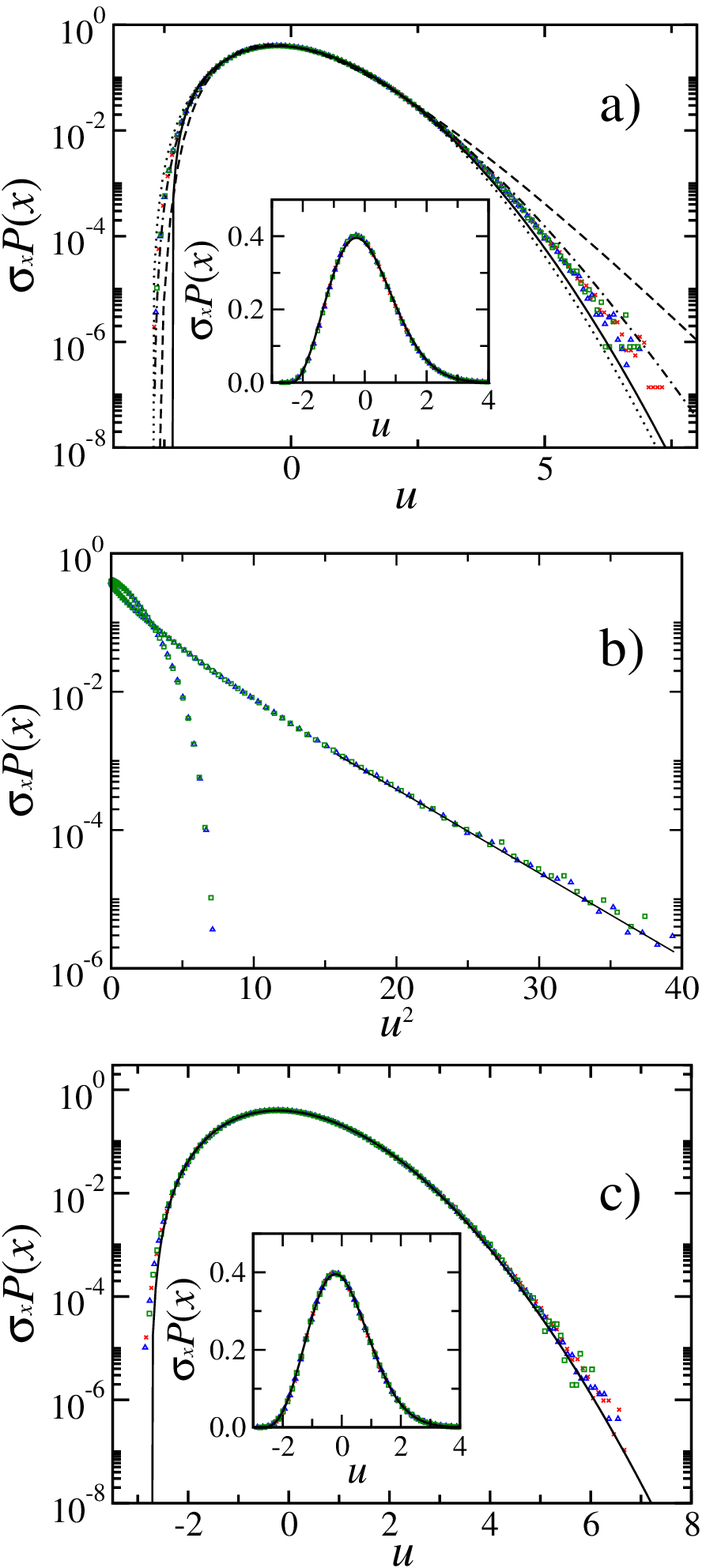}
\caption{(Color online) (a) Scaled CZD for point islands  with $i=1$ and $\theta=0.10$
[$R=10^{7}$ (crosses, red), $R=10^{8}$ (triangles, blue), and $R=10^{9}$ (squares, green)]
compared with the WS with $\beta=2$ (solid curve), the Gamma distribution (dashed curve),
the WS with $\beta=3$ (dotted curve), and the GG fit with $\beta=4$ and $n=1.5$ (dashed-dotted curve).
The inset is a linear-linear plot of the same data and the WS with $\beta=2$.
(b) Same simulation data of (a) with abscissa $u^2$. The solid straight line
highlights the Gaussian tail.
(c) Scaled CZD for point islands with $i=2$ and the same values of $\theta$ and $R$,
compared with the WS with $\beta=3$ (solid curve).
}
\label{fig1}
\end{figure}

\begin{figure}[!h]
\includegraphics[width=8cm]{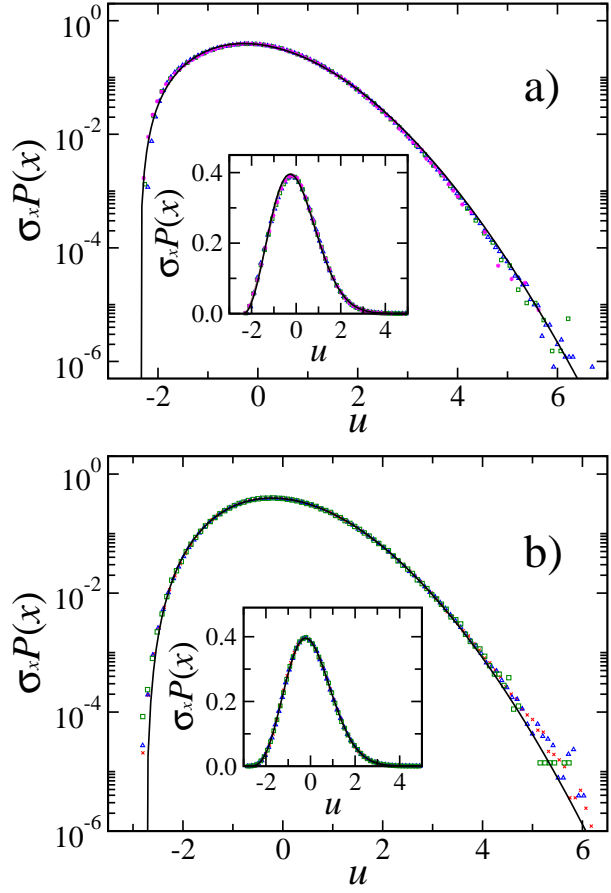}
\caption{(Color online) Scaled CZD for (a) fractal island model with $i=1$ and
(b) square island model with $i=2$, with $R=10^{7}$ (crosses, red),
$R=10^{8}$ (triangles, blue), $R=10^{9}$ (squares, green) and $R=10^{10}$ (stars, magenta).
The solid curves are the WS with  (a) $\beta=2$ and (b) $\beta=3$.
The insets are linear-linear plots of the same data.}
\label{fig2}
\end{figure}

\begin{figure}[!h]
\includegraphics[width=8cm]{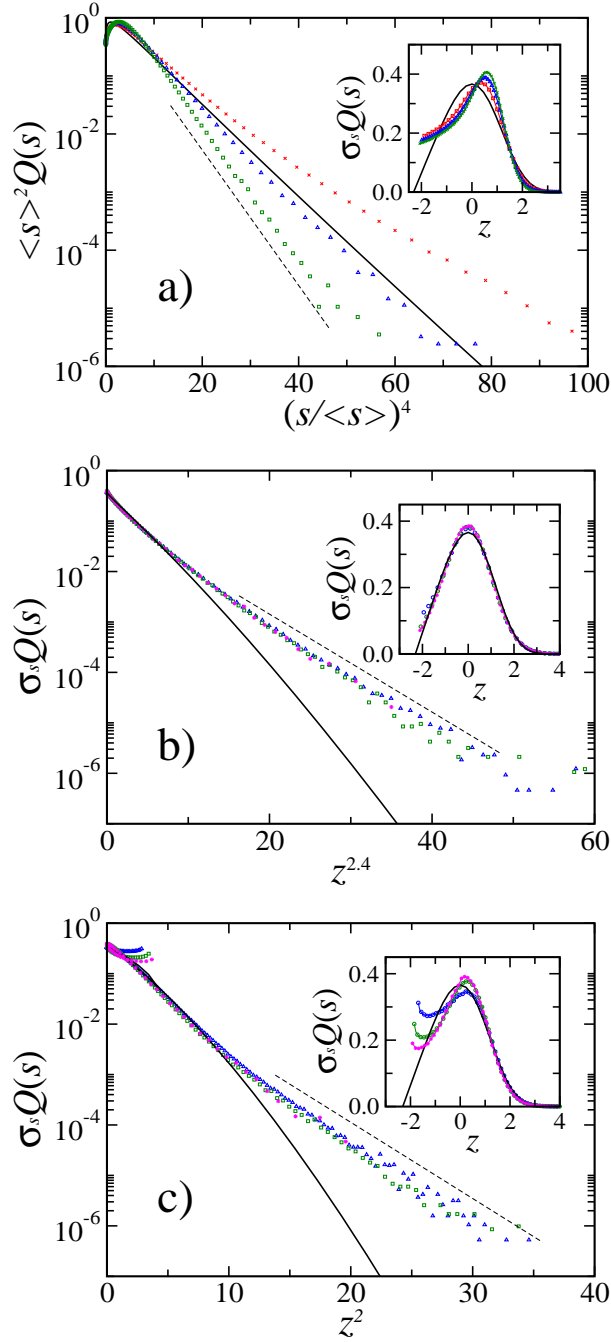}
\caption{(Color online) Scaled ISD for (a) point islands, (b) fractal islands, and (c)
square islands, with $i=1$ and $\theta=0.10$. Symbols correspond to $R=10^{7}$ (crosses, red),
$R=10^{8}$ (triangles, blue), $R=10^{9}$ (squares, green) and $R=10^{10}$ (stars, magenta).
The insets show linear-linear plots of the same quantities. The solid curve in all plots is the AF
formula with $i=1$. The dashed straight lines in the main plots are guides for the eyes.
}
\label{fig3}
\end{figure}

\end{document}